\documentclass{emulateapj}

\shorttitle{On the evolution of the molecular line profiles in C-shocks}
\shortauthors{I. Jim\'enez-Serra et al.}

\usepackage{multirow}
\begin{document}

\title{On the evolution of the molecular line profiles induced by 
the propagation of C-shock waves.} 

\author{I. Jim\'{e}nez-Serra\altaffilmark{1},
J. Mart\'{\i}n-Pintado\altaffilmark{2}, P. Caselli\altaffilmark{1}, 
S. Viti\altaffilmark{3} and A. Rodr\'{\i}guez-Franco\altaffilmark{2,4}}

\altaffiltext{1}{School of Physics \& Astronomy, E.C. Stoner Building,
The University of Leeds, Leeds, LS2 9JT, UK; I.Jimenez-Serra@leeds.ac.uk,
P.Caselli@leeds.ac.uk}

\altaffiltext{2}{Centro de Astrobiolog\'{\i}a (CSIC/INTA),
Instituto Nacional de T\'ecnica Aeroespacial,
Ctra. de Torrej\'on a Ajalvir km 4,
E-28850 Torrej\'on de Ardoz, Madrid, Spain; 
jmartin.pintado@iem.cfmac.csic.es,
arturo@damir.iem.csic.es}

\altaffiltext{3}{Department of Physics and Astronomy, University 
College London, WC1E$\,$6BT, London, UK; sv@star.ucl.ac.uk}

\altaffiltext{4}{Escuela Universitaria de \'Optica,  
Departamento de Matem\'atica Aplicada (Biomatem\'atica),
Universidad Complutense de Madrid,
Avda. Arcos de Jal\'on s/n, E-28037 Madrid, Spain}

\begin{abstract}

We present the first results of the expected variations of the molecular
line emission arising from material recently affected by C-shocks (shock 
precursors). Our parametric model of the structure of C-shocks has been 
coupled with a radiative transfer code to calculate the molecular 
excitation and line profiles
of shock tracers such as SiO, and of ion and neutral molecules 
such as H$^{13}$CO$^{+}$ and HN$^{13}$C, 
as the shock propagates through the unperturbed medium. 
Our results show that the SiO emission arising from the early stage 
of the magnetic precursor typically has 
very narrow line profiles slightly shifted in velocity with respect to the ambient 
cloud. This narrow emission is generated in the region where 
the bulk of the ion fluid has already slipped to larger velocities  
in the precursor as observed toward the young L1448-mm outflow. 
This strongly suggests that the detection of narrow SiO 
emission and of an ion 
enhancement in young shocks, is produced by the magnetic precursor of C-shocks.
In addition, our model shows that the different velocity components 
observed toward this outflow can be explained by the coexistence of 
different shocks at different evolutionary stages, within the same 
beam of the single-dish observations.    

\end{abstract}


\section{Introduction}

Shocks are an ubiquitous phenomena in the interstellar medium. 
In molecular dark clouds where the 
fractional ionisation of the gas is very low \citep[$\leq$10$^{-7}$; 
e.g.][]{cas98}, shock waves develop a 
magnetic precursor that accelerates, compresses and heats the ion 
fluid before the neutral one, forming a C-type shock 
\citep[see][]{dra80,flo95,flo96,les04}. 
The subsequent velocity decoupling between the ion and neutral components 
of the gas produces the sputtering of the mantles of dust grains, 
injecting large amounts of molecular material into the far downstream 
gas (e.g. Draine, Roberge \& Dalgarno 1983; Caselli, Hartquist  \& 
Havnes 1997; Schilke et al. 1997). 

Since silicon is heavily depleted onto dust grains in 
dark clouds such as L183 and L1448 \citep[SiO abundance of 
$\leq$10$^{-12}$;][]{ziu89,req07}, the detection of SiO is
a strong indicator of the dust grain processing by C-shocks. 
The SiO abundances in molecular outflows such as L1448-mm/IRS3 
are $\geq$10$^{-6}$, and the typical SiO emission arising from 
the warm postshock gas shows broad line profiles red- or blue-shifted 
with respect to the velocity of the ambient cloud 
\citep[e.g.][]{mar92}. 

\citet{lef98} and \citet{jim04} have reported the detection of very narrow 
SiO emission at ambient velocities with abundances of only
$\sim$10$^{-11}$-10$^{-10}$, toward the NGC1333 and L1448-mm molecular 
outflows. For L1448-mm, it has been proposed that this emission, which is 
also correlated with an enhancement of the ion and electron densities 
\citep{jim06}, arises from material recently processed by the magnetic 
precursor of C-shocks \citep{jim04}. 

Modelling of C-shock waves has traditionally focused on, and 
successfully reproduced,
the SiO abundances and line profiles generated in the evolved postshock gas 
\citep[][]{sch97,gus08}, but it has usually ignored that molecular abundances
and line profiles must change as C-shocks evolve in very young outflows 
like that powered by L1448-mm. \citet[][hereafter J08]{jim08} 
have shown that the observed trend for the enhancement of the SiO and 
CH$_3$OH abundances from the precursor to the high velocity components 
in this outflow \citep{jim05}, can be explained by the progressive 
sputtering of the grain mantles and cores as the shock evolves. 
However, no attempt has been made so far to predict the expected molecular
line profiles, and their changes, produced by the propagation of 
shock waves through the unperturbed gas.

Here we present the first results of the 
evolution of the line profiles of SiO, H$^{13}$CO$^{+}$ and HN$^{13}$C, 
produced by the propagation of C-shocks. The derived line profiles
strongly suggest that the narrow SiO lines and the ion enhancement are 
fingerprints of the magnetic precursor of shock waves. They also 
reproduce the broader moderate and high velocity emission components 
arising from the postshock gas in L1448-mm 
at later stages of the evolution of the shock. 

%FIG1**************************************
\begin{figure}
\epsscale{1.0}
\plotone{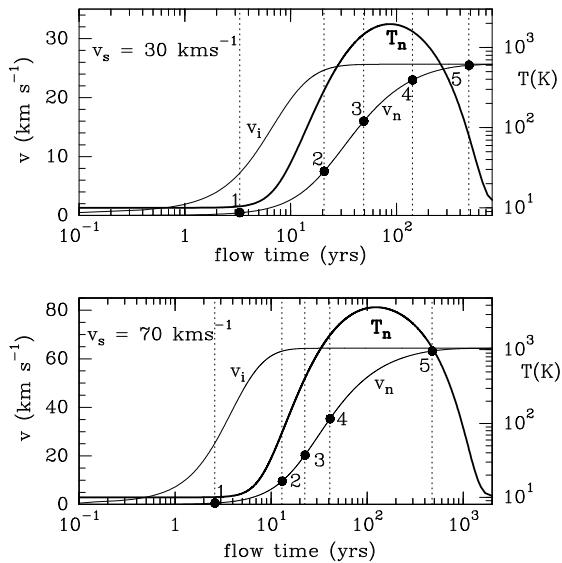}
\caption{Velocity of the ion and neutral fluids, $v_n$ and $v_i$, and 
temperature of the neutral gas, $T_n$, for two representative C-shocks
with $v_s$=30 and 70$\,$km$\,$s$^{-1}$. The initial H$_2$ density and magnetic
field are $n_0$=10$^5$$\,$cm$^{-3}$ and $B_0$=450$\,$$\mu G$.
The filled dots and vertical dotted lines 
indicate the flow times for which we 
{\it freeze} the evolution of the line profiles in both shocks 
(Sec.$\,$3 and Fig$\,$\ref{f2}). $T_n$ in our fit profile increases later in the
shock compared to MHD models. Although this is not crucial for the magnetic precursor
stage, it may significantly affect the derived SiO line profiles at stage 2 in the
70$\,$km$\,$s$^{-1}$ shock (see text in Sec.$\,$3 for details).}
\label{f1}
\end{figure}
%******************************************

\section{The model}

We consider a plane-parallel C-shock 
propagating with a velocity $v_s$ through the ambient medium.
The multi-layered steady-state physical structure of the shock in the 
frame of the preshock gas, is 
calculated by using the analytic approximation (fit) described in J08. 
Fig.$\,$\ref{f1} shows the shock field for $v_n$, $v_i$ and $T_n$, 
as a function of the flow time for two C-shocks with 
$v_s$=30 and 70$\,$km$\,$s$^{-1}$, initial H$_2$ density $n_0$=10$^5$$\,$cm$^{-3}$,
and magnetic field $B_0$=450$\,$$\mu G$ ($B_0$ is calculated following the relation 
$B_0$=$b_0$$\left(\frac{n_H}{cm^{-3}}\right)^{1/2}$ with $b_0$$\sim$1. 
See also Sec.$\,$2 in J08 for the 
details on the approximation). The Alfv\'en speed is 
$v_A$=2.18$\,$km$\,$s$^{-1}$ \citep[derived from $B_0$ and $n_0$ as $v_A$=$\frac{B_0}{\sqrt{4\pi m(H) n_0}}$; see][]{mye88}. 
This approximation has been validated in J08 (Sec.$\,$4 in this work) and  
reproduces the physical structure of C-shocks 
for $n_0$=10$^4$-10$^6$$\,$cm$^{-3}$. 
Hereafter, the term {\it time evolution} will be used for the variation 
of the molecular line emission as a function of the flow time, which
corresponds to the characteristic dynamical time across the shock. 

From Fig.$\,$\ref{f1}, it is clear that the ion fluid is accelerated before 
the neutrals by the magnetic precursor at the early stages 
of the shock. Since dust grains are charged particles, they are  
coupled to the magnetic field and move with the ion velocity, $v_i$. 
In the postshock gas, the velocities of the charged and neutral fluids re-couple 
again due to the transfer of momentum from the charged particles to the neutral fluid 
by collisions.

The line profiles of species linked to the grain chemistry 
such as SiO, are derived by considering the sputtering of the 
grain mantles and cores by collisions between dust grains and 
neutral particles such as H$_2$, He and other heavier species including CO. 
As shown by \citet{cas97} and J08, the abundances of sputtered material from dust grains,
and in particular of silicon, hardly change with the initial H$_2$ density of the gas. 
On the contrary, these abundances are drastically enhanced for increasing shock velocities due 
to the larger ion-neutral drift velocities, $|v_n-v_i|$, achieved in the shock.
We have assumed that a small fraction of Si ($\sim$0.01\% 
with respect to H$_2$O; J08) is present in the grain 
mantles. Projection effects are also taken into account in the calculation of the SiO 
abundances by including the inclination 
angle of the outflow with respect to the line of sight, $\theta$  
(J08). We consider that SiO is either directly released from 
the mantles \citep{jim05} or rapidly formed by the oxidation of 
Si sputtered from the cores 
\citep[$t$$\leq$25$\,$yrs;][]{cab07}. The gas phase abundance of SiO, 
$\chi$(SiO), is calculated as a function of the flow time, $t_n$, and the SiO 
column density in each plane-parallel slab of gas within the shock, $N$(SiO), 
is estimated as $N$(SiO)=$\chi$(SiO)$\,$$n$(H$_2$)$\,\Delta z$, where 
$n$(H$_2$) and $\Delta z$ are the H$_2$ density and spatial width of the slab, 
respectively. 

The level populations (i.e. excitation) of SiO in each slab of gas are 
calculated by using the LVG (Large Velocity Gradient) approximation. 
We use the H$_2$ collisional coefficients of SiO derived by
\citet{day06}. The line profiles are then obtained by making the radiative 
transfer through all individual slabs of material. 
The line peak optical depth, $\tau_0$, in each slab of gas is 
calculated by means of the LVG approximation through the following 
expression:

    \begin{eqnarray}
      \tau_0 & = & \left(\frac{c}{\nu_{ul}}\right)^3\frac{g_{up}}{\pi}\frac{A_{ul}}{dv_z/dz}\left(\frac{N_l}{g_l} - \frac{N_u}{g_u}\right) \label{opacity}
   \end{eqnarray}

\noindent
where $c$ is the speed of light, and $\nu_{ul}$ and $A_{ul}$ are the 
frequency and Einstein coefficient of the molecular transition 
for which we calculate the line profiles. $N_u$ and $N_l$ are the 
level populations of the upper and lower energy levels respectively, 
and $g_u$ and $g_l$, the degeneracy associated with them.  
$dv_z/dz$ is the radial velocity gradient along the propagation of the 
shock.     

To determine the behaviour of the optical depth, $\tau$, as a function
of velocity, we convolve the peak optical depth, $\tau_0$, at a given 
velocity (slab of material) with a Gaussian-like profile.
The linewidth of every individual profile, $\Delta v$, 
is estimated by assuming thermal and turbulent broadening, 
$\Delta v_{th}$ and $\Delta v_{tur}$. 
The thermal broadening depends on the temperature of 
the neutral fluid, $T_n$, as:

    \begin{eqnarray}
      \Delta v_{th} & = & \sqrt{\frac{8\,ln2\,k_B\,T_n}{m}} \label{vth}
   \end{eqnarray}

\noindent
where $k_B$ is the Boltzmann constant and $m$ is the mass of
the molecule. The thermal broadening for the SiO emission 
at the precursor stage with $T_n$=10$\,$K, is only of 
$\Delta v_{th}$=0.1$\,$km$\,$s$^{-1}$.

The turbulent linewidth is considered as a fixed parameter throughout 
the shock without largely affecting our estimates of the total linewidth.
Although turbulence in shocks still remains poorly understood, 
\citet{jou98} studied the effects of viscous dissipation and 
ion-neutral velocity decoupling on the chemistry and molecular line 
emission arising from turbulent flows (filamentary vortex) 
in the presence of magnetic fields in diffuse clouds.
In their model, \citet{jou98} determined that the maximum tangential 
velocity of the vortex filaments \citep[which gives the rms velocity 
dispersion of the large scale turbulence;][]{jim97,bel96}
 is close to the Alfv\'en velocity in the neutral fluid, $v_A$. 
If we therefore assume that $\Delta v_{tur}$ is directly 
proportional to $v_A$, the increase of $\Delta v_{tur}$ 
in the postshock gas can be estimated from the compression of 
the magnetic field and of the density of the gas at the late stages of the 
shock. This compression is given by the shock jump conditions 
$B$=$B_0$$(v_s/v_0)$ and $n$=$n_0$$(v_s/v_0)$, where $v_s$ is the shock speed 
and $v_0$ depends on $v_A$ and $v_s$ as described in J08. From here, 
$\Delta v_{tur}$ in the postshock gas ranges from $\sim$2.5 to 3.5 times 
larger than in the preshock gas for shock speeds between $v_s$=30 and 
70$\,$km$\,$s$^{-1}$. However, despite the increase by a factor
of $\sim$3 in $\Delta v_{tur}$, the total linewidth of the SiO line 
profile only differs by a factor of $\sim$1.2 from that calculated 
assuming constant turbulent broadening, because thermal broadening 
is dominant.   

For the ion and neutral species H$^{13}$CO$^{+}$ and HN$^{13}$C, 
we assume that their flux density, and therefore, their abundances
remain roughly constant across the shock. This is supported by 
the results of \citet[][]{mark00}, in which HCO$^{+}$ and HNC 
are chemically inert after the disruption of the grain mantles into the
gas phase by the passage of a C-shock. The initial column densities for
these molecular species are $N$$\sim$10$^{12}$$\,$cm$^{-2}$ \citep[][]{jim04},
and as the line profiles of 
H$^{13}$CO$^{+}$ and HN$^{13}$C get broader with the evolution of the shock, their 
peak column densities, $N_0$, are re-scaled 
following $N_0$=$N$$\,\delta v^*\,/\Delta v^*$, 
where $\delta v^*$ is the variation of the neutral/ion velocity 
from one slab of gas to the next one, and $\Delta v^*$ is the total
increase of the neutral/ion velocity from the beginning of the shock to the 
current slab of material in the shock. The excitation of H$^{13}$CO$^{+}$ 
and HN$^{13}$C has been derived by using the LVG approximation 
and the H$_2$ collisional coefficients of \citet{scho05}. 
The line profiles are derived following the same procedure as 
that used for SiO. Finally, we assume that the emission of 
H$^{13}$CO$^{+}$ and HN$^{13}$C, as well as that of SiO, has the same spatial 
extent and morphology. Beam filling factors
have not been considered in these calculations. The line intensities therefore depend
on the size of the shocked region.

A bow-shock morphology for the shock 
may lead to substantial variations on the derived SiO line profiles 
with respect to those calculated in the plane-parallel case. The planar geometry is 
however easier to handle in the calculation, and since the 
SiO bullets in L1448-mm only show a partial bow-shock morphology
\citep[see][]{dut97}, 
the use of both model 
geometries are equally justified.

\section{Time evolution of the molecular line profiles in C-shocks}

%FIG2**************************************
\begin{figure*}
\epsscale{.9}
\plotone{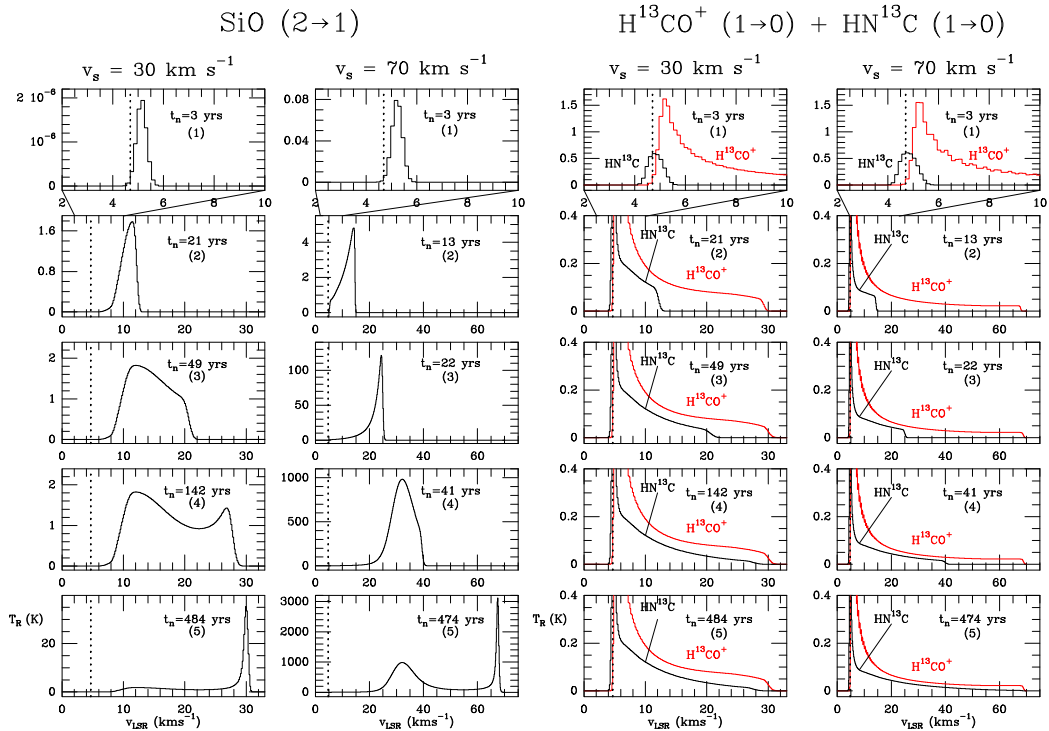}
\caption{Time evolution of the 
SiO 2$\rightarrow$1, the H$^{13}$CO$^{+}$ 1$\rightarrow$0 and  the 
HN$^{13}$C 1$\rightarrow$0 line profiles along the propagation of 
two C-shocks with $v_s$=30 and 70$\,$km$\,$s$^{-1}$. 
Numbers in parenthesis indicate the stages at which 
the profiles are {\it frozen} (see Fig.$\,$\ref{f1}). 
The flow times are shown in the upper part of all panels. 
Line intensities are in units of radiation temperature and the 
vertical dotted lines show the velocity of the preshock medium 
(4.7$\,$km$\,$s$^{-1}$).}
\label{f2}
\end{figure*}
%******************************************

In Fig.$\,$\ref{f2}, we show the evolution of the line profiles of SiO, 
H$^{13}$CO$^{+}$ and HN$^{13}$C obtained by {\it freezing} 
the propagation of two C-shocks, one with $v_s$=30$\,$km$\,$s$^{-1}$ and the 
other with $v_s$=70$\,$km$\,$s$^{-1}$, at 
5 different flow times (see dots in Fig.$\,$\ref{f1}). 
In order to directly compare
our results with observations, we have included the preshock gas 
velocity, $v_{cl}$, in the calculation of the molecular line profiles. 
For the L1448-mm outflow (Sec.$\,$4), we 
assume that $v_{cl}$=4.7$\,$km$\,$s$^{-1}$ 
and the inclination angle is $\theta$=70$^\circ$ \citep[][]{gir01}. 
The SiO, H$^{13}$CO$^{+}$ and HN$^{13}$C line profiles of Fig.$\,$\ref{f2} are
then derived as a function of the radial velocity measured 
by the observer, $v_{LSR}$. While for HN$^{13}$C this velocity is 
$v_{LSR}$=$v_{cl}+v_n$, for H$^{13}$CO$^{+}$ $v_{LSR}$ is 
$v_{cl}+v_i$. For SiO, one may think that this molecular
species would move with the ion fluid once it is ejected from grains. 
However, since SiO is not coupled to the magnetic 
field, this molecule is rapidly slowed down by collisions with H$_2$ 
to the velocities of the neutral fluid in time-scales, 
$t_s$, of only few days. For a SiO molecule 
$t^{-1}_s$=$n(H_2)\,\pi a(H_2)^2\sqrt{\frac{3\,k_B\,T_{n}}{m(H_2)}}$$\approx$2$\times$10$^{-12}$$T_n^{1/2}$$n_0$$\,$sec$^{-1}$. Assuming 
$n_0$=10$^5$$\,$cm$^{-3}$, $T_{n}$=10$\,$K and $a(H_2)$=0.7414$\,$\AA$\,$ as 
the internuclear distance for a H$_2$ molecule \citep[][]{hub79}, the slow down 
time is $t_s$$\sim$20$\,$days.

From Fig.$\,$\ref{f2}, we find that the SiO emission generated in 
shocks has very different line shapes depending on the evolutionary stage 
of the shock. In particular, it is striking that the typical SiO line 
profiles arising from the magnetic precursor at $t_n$=3$\,$yrs, have 
very narrow linewidths of only $\sim$0.5$\,$km$\,$s$^{-1}$ and are 
slightly shifted in velocity ($v_{LSR}\sim$5.2$\,$km$\,$s$^{-1}$) 
with respect to the ambient cloud ($v_{LSR}\sim$4.7$\,$km$\,$s$^{-1}$). 
The narrow linewidths of this emission are due to the low temperatures 
of the neutral fluid at the magnetic precursor stage 
($T_n$=10$\,$K; see Fig.$\,$\ref{f1}), and the small velocity shift 
is explained by the fact that the sputtering of the 
grain mantles requires ion-neutral drift velocities of
$|v_n-v_i|$$\geq$6$\,$km$\,$s$^{-1}$ to generate SiO abundances of 
$\geq$10$^{-12}$. This occurs when H$^{13}$CO$^{+}$ reaches velocities of 
$v_{LSR}$$\sim$12$\,$km$\,$s$^{-1}$ that imply  
$|v_n-v_i|$$\sim$7$\,$km$\,$s$^{-1}$ (see Fig.$\,$\ref{f1}). The peak 
intensity of the narrow SiO line is 4 orders of magnitude larger for the 
$v_s$=70$\,$km$\,$s$^{-1}$ shock (0.08$\,$K vs. 2$\times$10$^{-6}$$\,$K) due to the 
larger SiO abundances ($\sim$10$^{-8}$ vs. 10$^{-11}$) injected from 
the grain mantles into the gas phase. This intensity is increased by only a factor of 3 
if we consider that the fraction of silicon within the mantles is of 
0.1\% with respect to H$_2$O (i.e. 1 order of magnitude larger than that
assumed in Sec.$\,$2). This is due to the fact that only a fraction of the grain
mantles ($\leq$50\%) is released into the gas phase for the same flow time of 
$t_n$=3$\,$yrs.

For later flow times, the narrow SiO feature is swept out by
the larger efficiency of the sputtering of the grain cores once the velocity threshold of 
$\geq$25$\,$km$\,$s$^{-1}$ is overcome \citep[see e.g.][]{cas97,may00}. 
This enhances the gas phase abundances of SiO by several orders of magnitude. 
For the 30$\,$km$\,$s$^{-1}$
shock, these abundances are of $\sim$10$^{-8}$ and 10$^{-7}$ at $t_n$=10 and 
20$\,$yrs (see J08). For the 70$\,$km$\,$s$^{-1}$ shock, the SiO 
abundances are enhanced up to 5$\times$10$^{-6}$ and 7$\times$10$^{-6}$ at
$t_n$=13 and 22$\,$yrs, respectively. The increase of $T_n$ 
with the propagation of the shock leads to larger peak 
intensities and broader SiO line profiles. The SiO emission subsequently 
develops double peaked line profiles. The lower velocity peak is centred at  
$v_{LSR}\sim$12$\,$km$\,$s$^{-1}$ for $v_s$=30$\,$km$\,$s$^{-1}$ and 
at 32$\,$km$\,$s$^{-1}$ for $v_s$=70$\,$km$\,$s$^{-1}$. This peak is correlated with the 
maximum ejection of silicon from the grain cores ($\sim$10$^{-7}$ and 
7$\times$10$^{-6}$, i.e. 0.3\% and 19\% of all silicon locked in the cores for a 
30$\,$km$\,$s$^{-1}$- and a 70$\,$km$\,$s$^{-1}$-shock, 
respectively), which happens as soon as the ion-neutral drift velocity
reaches the threshold value of 25$\,$km$\,$s$^{-1}$. 
Since the fraction of silicon released into the gas phase is 
larger for the 70$\,$km$\,$s$^{-1}$ shock (by almost a factor of 100), the 
peak intensities therefore differ by more than 2 orders of magnitude 
from the $v_s$=30$\,$km$\,$s$^{-1}$ shock. 
We note that the moderate velocity peak was also present in the \citet{sch97} 
and \citet{gus08} models, but averaged with and dominated by the brightest
high velocity postshock emission. These moderate velocity profiles are similar 
in shape to those reported by \citet{cab07} for the young HH212 SiO microjet
and by \citet{lef98} in the NGC1333 outflows. 

The high velocity peaks at $v_{LSR}\sim$30 and 70$\,$km$\,$s$^{-1}$ 
($t_n$$\sim$500$\,$yrs; Fig.$\,$\ref{f2}), are produced
by the large accumulation of SiO in the postshock wide layers (slabs) of gas. 
Although the derived brightness temperatures are very high, this 
velocity component qualitatively resembles the SiO line profiles
obtained by \citet{sch97} and \citet{gus08}, 
and those observed by \citet{jim04} and \citet{nis07}. The lack of 
SiO gas-phase chemistry or depletion onto grains in our model may account for the
too strong emission predicted to arise from the high velocity postshock gas 
(see Sec.$\,$4).  

We note that discrepancies in $T_n$ do exist between the fit profile 
calculated with our approximation and the solutions 
derived from comprehensive MHD models where $T_n$ increases earlier in 
the shock \citep[e.g.][see also Fig.$\,$2 in J08]{kau96}. 
This is not crucial for the precursor stage where MHD calculations predict an 
initial low-temperature {\it plateau} for $T_n$ 
\citep[with $T_n$$\sim$10-20$\,$K; see e.g.][]{flo03}. However, 
for the moderate and high velocity components,  
the earlier increase of $T_n$ in the shock may produce significant variations 
in the peak intensity of the SiO line profiles compared to our results. 
While for the 30$\,$km$\,$s$^{-1}$ shock the SiO peak intensity could be increased
by only a factor of 2, for the 70$\,$km$\,$s$^{-1}$ shock this intensity
could be {\it enhanced} by more than 1 order of magnitude. 
Therefore, although {\it quantitative} results cannot be given for the moderate and
high velocity components, specially for the 70$\,$km$\,$s$^{-1}$ shock 
(also note that out model lacks of SiO gas phase chemistry and freeze-out; see Sec.$\,$4
for a more detailed discussion), the {\it qualitative} behaviour 
of the SiO line profiles is preserved.  
Quantitative predictions of the SiO line fluxes across the shock
will be carried out in the future with a comprehensive MHD code including 
chemical processes. 

Regarding the broadening of the SiO lines, the earlier increase of $T_n$ 
in the shock does not substantially affect the derived 
linewidths of the SiO line emission. Indeed, for both cases with
$v_s$=30$\,$km$\,$s$^{-1}$ and $v_s$=70$\,$km$\,$s$^{-1}$, these linewidths differ by less 
than a factor of 1.4 from those shown in Figure$\,$$\ref{f2}$ at
every evolutionary stage within the shock.

For H$^{13}$CO$^{+}$ and HN$^{13}$C, the differences in the line profiles of 
these species become apparent at the magnetic 
precursor stage. The line profile of HN$^{13}$C peaks at the preshock velocity of 
4.7$\,$km$\,$s$^{-1}$ at $t_n$=3$\,$yrs (stage 1),
and has a linewidth of $\sim$0.7-0.8$\,$km$\,$s$^{-1}$. The line 
emission of H$^{13}$CO$^{+}$, however, is (red)shifted to 
$v_{LSR}$=5.2$\,$km$\,$s$^{-1}$ and shows a moderate velocity wing 
with a terminal velocity (maximum velocity attained by the ion fluid at that 
certain flow time and reproduced in the molecular line profile) 
of $\sim$12$\,$km$\,$s$^{-1}$ for 
$v_s$=30$\,$km$\,$s$^{-1}$, and of $\sim$23$\,$km$\,$s$^{-1}$ for
$v_s$=70$\,$km$\,$s$^{-1}$ (see Fig.$\,$\ref{f1}). 
We note that the shift in velocity of the H$^{13}$CO$^{+}$ intensity peak is 
sensitive to the time step used in the calculation of the profiles. 
The line profiles of H$^{13}$CO$^{+}$ 
and HN$^{13}$C get broader as the shock evolves. While the terminal velocity 
of the H$^{13}$CO$^{+}$ emission is reached at stage 2 ($t_n$=21 and 13$\,$yrs; 
see Fig.$\,$\ref{f2}), the line profiles of 
HN$^{13}$C require flow times of $t_n$=500$\,$yrs to show broad 
emission with a similar terminal velocity.   

\section{Observational fingerprints of the magnetic precursor of C-shocks}

%FIG3**************************************
\begin{figure*}
\begin{center}
\includegraphics[angle=270,width=0.8\textwidth]{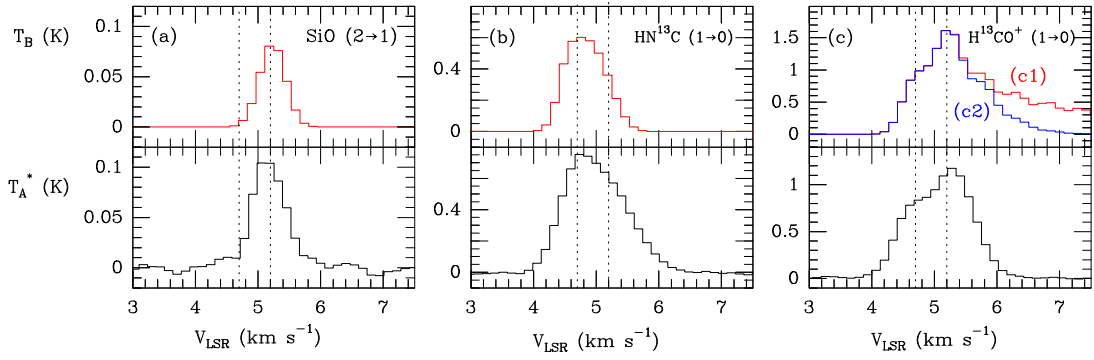}
\caption{Comparison between the predicted line profiles of SiO, HN$^{13}$C and 
H$^{13}$CO$^{+}$ at $t_n$=3$\,$yrs for a shock with $v_s$=70$\,$km$\,$s$^{-1}$ and 
$n_0$=10$^5$$\,$cm$^{-3}$ (upper panels), and the line 
profiles observed toward the L1448-mm outflow \citep[lower panels;][]{jim04}.  
The ambient (4.7$\,$km$\,$s$^{-1}$) and the precursor (5.2$\,$km$\,$s$^{-1}$) 
components are indicated by the vertical dotted lines. In the upper
(c) panel, we show the line profiles of H$^{13}$CO$^{+}$ calculated 
with and without ion recombination (blue c2 and red c1 
lines, respectively).}
\end{center} 
\label{f3}
\end{figure*}
%******************************************

%FIG4**************************************
\begin{figure}
\epsscale{0.9}
\plotone{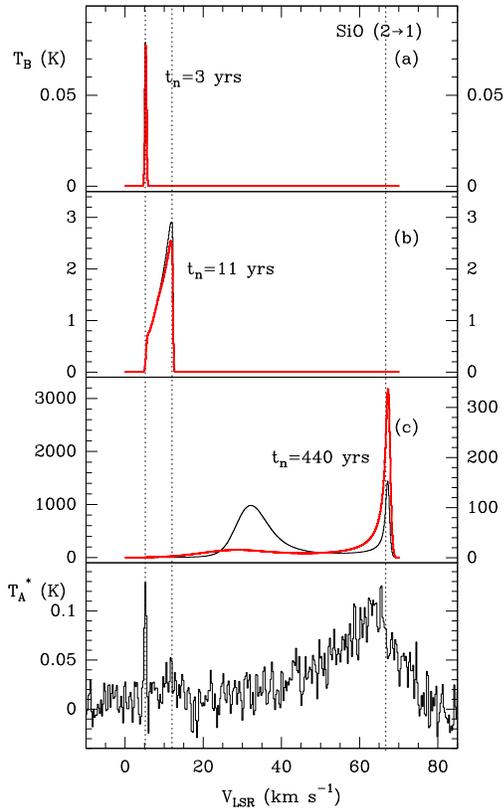}
\caption{SiO 2$\rightarrow$1 line profiles (black solid lines) 
derived for the same C-shock model  
($v_s$=70$\,$km$\,$s$^{-1}$ and $n_0$=10$^{5}$$\,$cm$^{-3}$) shown in Figs.$\,$\ref{f2}
and \ref{f3}, which best fit the full observed SiO line profile toward L1448-mm
\citep[lower panel;][]{jim04}. 
The freezing flow times ($t_n$=3, 11 and 440$\,$yrs) are shown in every panel.  
The vertical dotted lines indicate the radial velocities ($v_{LSR}$=5.2,
12 and 68$\,$km$\,$s$^{-1}$) that correspond to these flow 
times. We also include the results of a second C-shock model that considers the 
possible effects that gas-phase chemistry 
may have on the derived SiO line profiles (red thick solid lines).} 
\label{f4}
\end{figure}
%******************************************

In panels (a), (b) and (c) of Fig.$\,$\ref{f3}, we show the comparison 
between the line emission of SiO, HN$^{13}$C and H$^{13}$CO$^{+}$ observed 
toward the L1448-mm outflow \citep{jim04}, and the line profiles 
obtained by our model for a C-shock with $v_s$=70$\,$km$\,$s$^{-1}$, 
$n_0$=10$^5$$\,$cm$^{-3}$, $B_0$=450$\,$$\mu G$, 
$T_0$=10$\,$K and $v_{cl}$=4.7$\,$km$\,$s$^{-1}$ at the magnetic 
precursor stage with $t_n$=3$\,$yrs. The shock velocity of 
70$\,$km$\,$s$^{-1}$ is selected based on the terminal 
velocity of the SiO emission in L1448-mm. Although the 
time-scales of 3$\,$yrs at the precursor are 
very short, they are consistent with the dynamical ages  
of the very young shocks detected in the vicinity of the L1448-mm 
source \citep[$t$$\sim$90$\,$yrs;][]{gir01},
and in the HH212 SiO microjet \citep[$t$$\leq$25$\,$yrs;][]{cab07}. 
The freezing flow time of $t_n$=3$\,$yrs implies a radial velocity of 
$v_{LSR}$=5.2$\,$km$\,$s$^{-1}$ (i.e. $v_n$=0.5$\,$km$\,$s$^{-1}$), 
at which the observed 
narrow SiO lines show their emission peaks toward L1448-mm. The predicted 
peak intensity for the $J$=2$\rightarrow$1 transition is of 0.08$\,$K and
implies a SiO abundance of 2$\times$10$^{-8}$. We note that we require
a shock with $v_s$=70$\,$km$\,$s$^{-1}$ in order to inject this SiO abundance
at this early stage of the shock. One may think that SiO abundances of $\sim$10$^{-8}$ 
could also be obtained
at later flow times in the shock. However, the calculated line profile would
be considerably broader and its emission peaks would be largely (red)shifted, which
is not consistent with the narrow SiO emission observed toward this outflow. Therefore, 
the radial velocity and the narrow linewidth of the narrow SiO emission are
strong constraints for the determination of the shock age associated with this
emission.

The derived peak intensity of the narrow SiO emission shown in Fig.$\,$\ref{f3} only
differs by 20\% from the observed one. The estimated
$J$=3$\rightarrow$2/$J$=2$\rightarrow$1 line intensity ratio (of $\sim$0.4)
is, however, a factor of 3 smaller than that measured in L1448-mm 
\citep[see][]{jim04}, suggesting that the H$_2$ densities toward this 
outflow might be larger than those considered in our model.  
%The lack of detection of the $J$=5$\rightarrow$4 
%line emission (3$\sigma$ noise level of $\leq$0.171$\,$K) by 
%\citet{jim06} is consistent with the peak intensity of only 
%6$\times$10$^{-4}$$\,$K derived by our model for this transition. 
The low velocity resolution (of $\geq$1$\,$km$\,$s$^{-1}$ vs. 
0.14$\,$km$\,$s$^{-1}$ shown in Fig.$\,$\ref{f3}) of the \citet{nis07}
observations toward this outflow, prevents the direct comparison of these 
single-dish data with our results for high-$J$ SiO transitions. 
In any case, the derived line intensities of the SiO transitions with $J_{up}$$\geq$5
are expected to be far below the 3$\sigma$ noise level of these 
observations.

The SiO gas phase abundance required to match the intensity of 
all narrow SiO lines is of $\sim$2$\times$10$^{-8}$, three 
orders of magnitude larger than that derived by \citet{jim04}. Beam 
dilution could explain the discrepancies in the
SiO abundances. If the emitting region of the narrow SiO 
$J$=2$\rightarrow$1 emission is 
$\leq$1$''$ (of $\sim$1.5-3$\times$10$^{15}$$\,$cm for the magnetic precursor 
at the distance of 200$\,$pc; see J08), 
the predicted abundance of $\sim$2$\times$10$^{-8}$ is diluted to 
$\sim$10$^{-11}$, which is of the same order of magnitude as that 
measured in this outflow \citep{jim04}. 

From Fig.$\,$\ref{f3}, we also find that the difference between  
the H$^{13}$CO$^{+}$ and the HN$^{13}$C line profiles observed by 
\citet{jim04} toward L1448-mm, are reproduced by the model. While the 
neutrals (HN$^{13}$C; panel b) 
remain unperturbed at the early stages of the shock (their line profiles are 
centred at the velocity of the preshock gas of 4.7$\,$km$\,$s$^{-1}$),
the ion H$^{13}$CO$^{+}$ fluid (panel c) has slipped to larger 
velocities (with its line peak centred at 
$\sim$5.2$\,$km$\,$s$^{-1}$) and shows an ion enhancement 
in the precursor component that is consistent with that measured in 
L1448-mm \citep{jim04}. 

In contrast with the observed line profile of H$^{13}$CO$^{+}$, 
however, the calculated line of this species shows redshifted 
broad emission with terminal velocities of $\leq$12$\,$km$\,$s$^{-1}$ 
(red c1 histogram; Fig.$\,$\ref{f3}). The observational procedure, 
frequency-switched mode, could be the responsible of the missing 
broad emission in the final spectrum of H$^{13}$CO$^{+}$. 
Another possibility could be that H$^{13}$CO$^{+}$ rapidly recombines  
in the magnetic precursor due to the large electron density 
enhancement at this stage of the evolution of the shock \citep[][]{flo03}. 
If we assume that the H$^{13}$CO$^{+}$ abundance  
decreases by ion recombination with the flow time 
\citep[we consider a fractional ionisation of 
$\sim$5$\times$10$^{-5}$ in the precursor 
component;][]{jim06}, the broad H$^{13}$CO$^{+}$ wing
emission vanishes, leading to a narrower H$^{13}$CO$^{+}$ line 
profile that mimics the observed one 
(terminal velocity of $\sim$7$\,$km$\,$s$^{-1}$; blue c2 histogram). 
%Future interferometric images of the HCO$^{+}$ and H$^{13}$CO$^{+}$ molecular
%ions will help to disentangle the origin of the missing broad H$^{13}$CO$^{+}$
%emission between these two possibilities. 
In the final derived spectrum of H$^{13}$CO$^{+}$, we have also 
included the ambient cloud component at 4.7$\,$km$\,$s$^{-1}$ with a 
H$^{13}$CO$^{+}$ peak column density of 
5$\times$10$^{11}$$\,$cm$^{-2}$ \citep{jim04}. The calculated line intensities
of H$^{13}$CO$^{+}$ and HN$^{13}$C differ by 13-50\% from 
those measured toward L1448-mm. Variations by a factor of 2 in the 
molecular column densities and initial H$_2$ density would lead to
H$^{13}$CO$^{+}$ and HN$^{13}$C line intensities that differ by less 
than a factor of 2 from the ones calculated for 
$N$(H$^{13}$CO$^{+}$)=$N$(HN$^{13}$C)=10$^{12}$$\,$cm$^{-2}$ and $n_0$=10$^{5}$$\,$cm$^{-3}$.
Interferometric observations that provide the morphology and spatial extent 
of this emission are therefore required to solve the discrepancies in both 
the molecular abundances and line peak intensities of H$^{13}$CO$^{+}$ and HN$^{13}$C
in the magnetic precursor.

\citet{jim04} discussed the possibility that narrow SiO could be 
generated by the evaporation of the icy grain mantles by the illumination of UV 
radiation in Photon Dominated Regions, or PDRs \citep{walm99}. 
\citet{sch01} reported the detection of narrow SiO emission
with SiO abundances of $\sim$10$^{-11}$ toward a sample of PDRs such as the 
Orion Bar or S 140. The linewidth of this emission is however of 
$\sim$2-3$\,$km$\,$s$^{-1}$, a factor of $\sim$5 broader than those 
measured toward L1448-mm. Furthermore, HCO, one of the main products of 
the UV photo-chemistry \citep[see e.g.][]{sche88,ger08} is not detected in the 
precursor component of this outflow, which allows to rule out the PDR origin for 
the narrow SiO emission. 

Alternatively, narrow SiO emission may arise from an intervening cloud where
turbulence motions of at least 6$\,$km$\,$s$^{-1}$ would be efficient enough
to erode the grain mantles and inject {\it detectable} abundances of SiO 
into the gas phase. However, in contrast with the large-scale
morphology expected from this scenario, the narrow SiO emission
is not widespread around L1448-mm as shown by the low upper limits of the
SiO abundance of $\leq$10$^{-12}$ found in the regions with no precursor interaction
\citep[][]{jim04,jim05}. We note that these upper limits are of the same order of 
magnitude as those measured in the quiescent gas of the L183 dark cloud 
by the high-sensitivity 30$\,$m observations of \citet[][]{req07}. This
indicates that the lack of SiO emission toward these regions is caused by a 
real lack of SiO in gas phase. From all this, we propose that the most
likely explanation for the narrow SiO lines detected toward L1448-mm, 
is the interaction of the magnetic precursor of C-shocks in very young 
outflows.

The line profiles calculated across our parametric C-shock 
structure also suggest that the moderate and high velocity SiO components
observed toward L1448-mm are linked to later stages in the evolution of
the shock. In Fig.$\,$\ref{f4}, we show the comparison between the full 
SiO line profile in L1448-mm (lower panel), and the SiO line emission 
predicted for a shock with $v_s$=70$\,$km$\,$s$^{-1}$ 
and $n_0$=10$^5$$\,$cm$^{-3}$ at three different evolutionary 
stages (black solid lines). The narrow SiO emission is exactly the same as that 
shown in panel (a) of Fig.$\,$\ref{f3} for $t_n$=3$\,$yrs, and the moderate velocity 
component is reproduced by a C-shock at $t_n$=11$\,$yrs. We note 
that the discrete nature of these two components is a direct consequence of the 
assumed chemical structure of dust grains: icy mantles surrounding 
the silicate/graphite cores,
which have different binding energies \citep[of 0.53$\,$eV and 50$\,$eV for the mantles and
the cores, respectively; see][]{tie94,may00}. The high velocity gas 
emission is matched with a C-shock at $t_n$=440$\,$yrs.

Although the model 
fails to reproduce the peak intensity of the moderate and high velocity 
SiO components, the line shapes are {\it qualitatively} similar to those observed in L1448-mm. 
At $t_n$=440$\,$yrs, however, the moderate velocity emission shows an excess in intensity 
which is difficult to match with the observations. The facts that depletion of Si/SiO 
onto dust grains and/or that silicon gas-phase chemistry have not been taken into account 
in our model, could explain these discrepancies. Including the depletion rate 
of \citet[][see Eq.$\,$4 in this work]{tie85} with a sticking coefficient of 
$\sim$1, the depletion of SiO onto dust grains reduces 
the abundance of this molecule by a factor of $\sim$1.2 for intermediate flow 
times and by a factor of $\sim$5 in the far postshock gas. 
More dramatically, the conversion of silicon into SiO via gas-phase 
reactions with OH and O$_2$ (e.g. Si+OH$\,$$\rightarrow$$\,$SiO+H and 
Si+O$_2$$\,$$\rightarrow$$\,$SiO+O) 
is known to be not 100\% efficient for high densities (of 
$n_0$=10$^{5}$-10$^6$$\,$cm$^{-3}$), for which OH and O$_2$ are partially destroyed 
by the atomic hydrogen generated in the
shock \citep{gus08}. This may lead to significantly lower SiO abundances \citep[by more 
than 1 order of magnitude; see][]{gus08} than those obtained directly from the 
sputtering of dust grains. 

If we now consider that the gas-phase abundance of 
SiO cannot exceed $\sim$10$^{-6}$ \citep[note that this is the maximum SiO abundance 
estimated in L1448-mm; see][]{mar92}, the SiO line profile derived at
$t_n$=440$\,$yrs {\it mimics} the emission observed toward this outflow (see red solid
line in panel (c) of Fig.$\,$\ref{f4}). The decrease of the gas-phase abundance of 
SiO does not largely affect the SiO lines calculated for 
the precursor and moderate velocity components at $t_n$=3 and 11$\,$yrs, because the 
grain core sputtering is not efficient enough to inject SiO abundances larger than 
10$^{-6}$ (see panels (a) and (b) in Fig.$\,$\ref{f4}). Beam filling factor 
corrections could also account for the different peak intensities obtained from
the model for the moderate and high velocity emission of SiO, assuming that 
both components show different spatial extents. High-angular resolution observations
with interferometers such as the PdBI and the SMA, are therefore strongly required not only 
to better constrain the initial conditions and the evolutionary stage of the different 
shocks suggested by the model, but also the chemical properties of the dust grains 
present in this star forming region.

Finally, shock velocities of 70$\,$km$\,$s$^{-1}$ 
clearly exceeds the critical velocities of C-shocks 
\citep[of $\sim$45-50$\,$km$\,$s$^{-1}$;][]{dra83}.
\citet{bou02} have shown that these critical velocities can 
be increased to few $\sim$100$\,$km$\,$s$^{-1}$ for moderate H$_2$ 
densities and high magnetic fields (of few $mG$) as those observed 
toward low mass star forming regions such as NGC1333 \citep{gir06}. 
However, we cannot rule out the possibility 
that the high velocity SiO emission toward L1448-mm is associated with 
a J-type discontinuity within the postshock gas in this outflow \citep{chi98}.

In summary, we have modelled for the first time the expected evolution 
of the line profiles of SiO, H$^{13}$CO$^{+}$ and HN$^{13}$C 
as C-shocks propagate through the ambient medium. The model suggests that the 
narrow SiO emission and the ion enhancement found in the young shocks of the
L1448-mm outflow, are signatures of the magnetic 
precursor of C-shocks. The different velocity components (precursor, 
moderate velocity and high velocity gas) observed toward this outflow could be
explained by the coexistence of C-shocks with 
$v_s$=70$\,$km$\,$s$^{-1}$ at different evolutionary stages 
within the single-dish beam of the SiO observations. 
The model will be used in the future to reproduce the molecular line 
emission observed toward a sample of molecular outflows under different 
physical conditions and different evolutionary stages.

\acknowledgments

We would like to thank the referee, Paul Ho, for his constructive comments
that largely improved the paper.
IJS and JMP acknowledges the Spanish MEC for the support provided 
through projects number ESP2004-00665, ESP2007-65812-C02-01 and
``Comunidad de Madrid'' Government under PRICIT project
S-0505$/$ESP-0277 (ASTROCAM).

\end{document}